\documentclass[12pt]{article}
\usepackage{amsmath,amssymb,amscd,cite}
\newtheorem{thm}{Theorem}[section]
\newtheorem{prop}[thm]{Proposition}
\newtheorem{lem}[thm]{Lemma}
\newtheorem{cor}[thm]{Corollary}

\newtheorem{defi}[thm]{Definition}
\newcommand{\pf}{{\bf Proof. \ }}
\newcommand{\qed}{\hfill $\Box$ \\}
\font\msbm=msbm10 at 12pt
\newcommand{\Z}{\mbox{\msbm Z}}

\newcommand{\F}{\mbox{\msbm F}}
\newtheorem{rem}[thm]{Remark}
\newtheorem{ex}[thm]{Example}

\newcommand{\ord}{ord}
\begin{document}

\title{On the Automorphism Groups and Equivalence of Cyclic Combinatorial Objects}

\author{ Kenza Guenda and T. Aaron Gulliver
%\date{}
\thanks{ T. Aaron Gulliver is
with the Department of Electrical and Computer Engineering,
University of Victoria, PO Box 3055, STN CSC, Victoria, BC, Canada
V8W 3P6. email: agullive@ece.uvic.ca.}}
\date{}
\maketitle

\begin{abstract}
We determine the permutation groups that arise as the automorphism
groups of cyclic combinatorial objects. As special cases we classify
the automorphism groups of cyclic codes. We also give the
permutations by which two cyclic combinatorial objects on $p^m$
elements are equivalent.
\end{abstract}
\section{Introduction}
Let $n$ be a positive integer and $S_n$ the group of permutations on $n$ elements.
A combinatorial object $\mathcal{C}$ on $n$ elements
is called cyclic if its automorphism group $Aut(\mathcal{C})$
contains the complete cycle $T=(1,2,\ldots,n)$ of length $n$.
Let $C_p$ denote the cyclic group of order $p$.

The class of cyclic objects includes circulant graphs, circulant digraphs, cyclic designs
and cyclic codes. Two cyclic combinatorial objects $\mathcal{C}$ and
$\mathcal{C}'$ on $n$ elements in the same category of objects are
said to be equivalent if there exists a permutation $\sigma$ of the
symmetric group $S_n$ acting on $\{0,1,\ldots, n-1\}$ such that $\mathcal{C}'=\sigma \mathcal{C}$.
When $n$ is a prime number, it has
been proven by Bays and Lambossey~\cite{bay,lambossey} that circulant
graphs are equivalent by a permutation $\mu$ if and only if $\mu$
satisfies $\mu(i)=ai \pmod n$ for $a \bmod n$ such that $(a,n)=1$.
Such permutations are called multipliers.
Alspach and Parson~\cite{alspach} proved that two circulant graphs or digraphs on
$pq$ vertices, where $p>q$ are two distinct primes, are equivalent if
and only if they are equivalent by a multiplier.
These results hold for all cyclic combinatorial objects in the case that
$Aut(\mathcal{C})$ has a $p$-Sylow subgroup of order $p$~\cite[Case 1]{alspach}.
Muzychuk~\cite{mike} proved that this result also holds for circulant graphs in the square-free case.
More generally, Palfy proved that if $n$ is such that $(n,\phi(n))=1$, or
$n=4$, then two cyclic combinatorial objects are equivalent if and only if they are equivalent by a multiplier.
Furthermore, Palfy gave a class of cyclic combinatorial objects on $n$ elements where $n\neq
4$ and $\gcd(n,\phi(n))\neq 1$ such that the class contains equivalent objects that are not equivalent by a multiplier.
% The problem of isomorphism for circulant graphs of length of length $p^n$ has been
%completely solved by Klin and P\:oschel~\cite{klin}.
% Furthermore
%Klin proved~\cite{klin2} that under some conditions solving the
%problem of isomorphism of cyclic combinatorial objects on $n$
%elements is reduced to solve the problem of the combinatorial
%objects on a prime power.

Brand~\cite{brand} proved that cyclic combinatorial objects on $p^m$
elements can be equivalent only under a specific set of permutations
which depends on the $p$-Sylow subgroup of $Aut(\mathcal{C})$.
Building on Brand's results~\cite{brand}, Huffman et
al.~\cite{vanessa} solved the equivalency problem for cyclic
combinatorial objects and cyclic codes in the case $n=p^2$. This was
achieved by explicitly determining the set of permutations under
which two cyclic combinatorial objects or extended cyclic objects
can be equivalent. In~\cite{vanessa}, a negative answer was given to
the generalization of their results to the case $n=p^m$, $m>2$. This
is due to the fact that the polynomials of Brand that are crucial to
proving the results do not generate a Sylow subgroup of
$S_{p^m}$. More recently, Babai et al.~\cite{babai} gave an
exponential time algorithm for determining the equivalence of two
linear codes.
%One of the aims of this paper is explicitly give the set of permutations for
%cyclic combinatorial objects on $p^r$ elements.
In this paper, we consider the equivalency problem of cyclic
combinatorial objects of length $p^m$. We generalize the results
of~\cite{vanessa} (which are only for the case $n=p^2$), by
explicitly giving the permutations by which two cyclic codes of
length $p^m$ are equivalent. This allows us to develop an algorithm
which solves the equivalency problem by checking no more than
$\langle \log_2 (p-1) \rangle+1$ permutations in the automorphism
group. We also classify the automorphism groups of cyclic
combinatorial objects.
%
%The motivation for this is that the for our method of the equivalency problem we
This requires knowledge of the $p$-Sylow subgroup of $Aut(\mathcal{C})$.
We consider the special case of cyclic codes.
Even though these codes are well known and have been studied extensively, very little is known about their automorphism groups,
in particular the BCH and Reed-Solomon codes~\cite{berger91,berger96,berger96b}.
% The last situation implies that the code is trivial.

Beside the theoretical interest in automorphism groups and
equivalence, there are many practical applications. Algorithms
capable of determining graph equivalency can be used in optical
character recognition~\cite{wag} and image processing. Further, the
automorphism groups and equivalency of cyclic codes can be employed
in permutation decoding~\cite{bhargava} and determining the weight
distribution of a code~\cite{macwilliams}. While the equivalency of
cyclic design find application in optical orthogonal
codes~\cite{cao}. The remainder of this paper is organized as
follows. In Section 2, we investigate the automorphisms of cyclic
objects. We also classify the automorphism groups of cyclic
combinatorial objects. Section 3 considers the automorphism groups
of cyclic codes. New results concerning the automorphism groups of
cyclic codes of length $p^m$ are presented. We also give an
algorithm to find cyclic codes of length $p$. This algorithm
requires that only $p-1$ permutations be checked. In Section 4, we
simplify and generalize some results of Huffman et
al.~\cite{vanessa} from length $p^2$ to length $p^m$. This allows us
to provide a solution to the equivalency problem for cyclic
combinatorial objects. An algorithm to solve the equivalency problem
is then presented which requires checking at most $\lceil \log_2
(p-1)\rceil +1$ permutations.

%Note that the automorphism group of codes can find application in
%finding minimum weight and decoding. While the equivalency problem
%for both codes and deign find many applications such as the problem
%o
Throughout this paper, $\ord_n(q)$ denotes the multiplicative
order of $q$ modulo $n$. In other words it is the smallest integer
$r$ such that $q^r\equiv 1 \pmod n$. The group $Aut(\mathcal{C})$
denotes the automorphism group of the object $\mathcal{C}$ (with
elements which are permutations from $S_n$). We denote by $z$ the
largest integer such that $p^z | (q^{t'}-1)$, where $t'$ is the
order of $q$ modulo $p$.

\section{The Automorphism Groups of Cyclic Objects}
We begin this section with some well known definitions.
Let $n$ be a positive integer.
The set of permutations
$AG(n)=\{\tau _{a,b} : a \neq 0, (a,n)=1, b \in \mathbb{Z}_n\}$ is
the subgroup of $S_n$ formed by the permutations defined as follows
\begin{equation}
\begin{split}
\tau _{a,b} :\mathbb{Z}_n & \longrightarrow  \mathbb{Z}_n \\
          x &\longmapsto (ax +b )\bmod n.
\end{split}
\end{equation}
The group $AG(n)$ is called the group of affine transformations.
The affine transformations $M_a=\tau_{a,0}$ are also multipliers.
The affine group $AGL(1,p)$ is the group of affine transformations over $\mathbb{Z}_p$.
The projective semi-linear group $P\Gamma L(d,t)$ is the semi-direct product of the projective linear group $PGL(d,t)$
and the automorphism group $Z=Gal(\mathbb{F}_t/\mathbb{F}_{p})$ of
$\mathbb{F}_t$, where $t=p^s$, $p$ prime, i.e.
\[
P\Gamma L(d,t)= PGL(d,t)\rtimes Z.
\]
The orders of these groups are $|PGL(d,t)|=(d,t-1)|PSL(d,t)|, |Z|=s$
and $|P\Gamma L(d,t)|=s|PGL(d,t)|$.
\begin{rem}
\label{rem:iso} If $(d,t-1)=1$, then $PGL(d,t)=PSL(d,t)$. If $t$ is
a prime we have $ P\Gamma L(d,t)= PGL(d,t)$.\qed
\end{rem}

From the fact that the automorphism group of a cyclic combinatorial
object contains the complete cycle $T$ of length $n$, we can easily prove that this group is transitive.
A transitive group is either primitive or imprimitive.
An interesting class of primitive groups is the class of doubly-transitive groups.
A doubly-transitive group $G$ has a unique minimal normal subgroup $N$ which is either
regular and elementary abelian or simple and primitive, and has centralizer $C_G(N)=1$~\cite[p. 202]{burnside}.
All simple groups which can occur as a minimal normal subgroup of a doubly-transitive group are known.
This result is due to the classification of finite simple groups.
Using this classification, McSorley~\cite{sorley} gave the following result.
\begin{lem}
\label{lem:sorley} A group $G$ of degree $n$ which is
doubly-transitive and contains a complete cycle has socle $N$ with
$N \leq G \leq Aut(N)$, and is equal to one of the cases in Table
1. \qed
\end{lem}
\begin{table}
\caption{The Doubly Transitive Groups that Contain a Complete Cycle}
\begin{center}
\begin{tabular}{|c|c|c|}
\hline
 $G$ & $n$ & $N$  \\
\hline
$AGL(1,p) $ & $p$ &$C_p$ \\
$S_4 $ & $4$ &$C_2 \times C_2$ \\
$S_n , n\geq 5$ & $n$ &$Alt(n)$ \\
$Alt(n),n \text{ odd and } \geq 5$ & $n$ &$Alt(n)$ \\
 $PGL(d,t) \leq G \leq P\Gamma L(d,t)$ & $\frac{t^d-1}{t-1}$&
 $PSL(d,t)$ \\
$(d,t)\neq  (2,2),(2,3),(2,4) $&&\\
$PSL(2,11)$& 11 &$PSL(2,11)$\\
$M_{11}$(Mathieu)&11&$M_{11}$(Mathieu)\\
$M_{23}$(Mathieu)&23&$M_{23}$(Mathieu)\\
\hline
\end{tabular}
\end{center}
\end{table}

\noindent
As a direct application of Lemma~\ref{lem:sorley}, we obtain the following result.
\begin{thm}
\label{lem:bur} Let $\mathcal{C}$ be a cyclic combinatorial  objects
on $p$ elements. Then $Aut(\mathcal{C})$ is a primitive group
with socle $S$, and one of the following holds:
\begin{itemize}
\item[(i)] $Aut(\mathcal{C})=S_n$ or $Alt(n)$.
\item[(ii)] $Aut(\mathcal{C})$ is a solvable group of order $pm$ with $m$ a divisor of $p-1$ and $S=C_p \leq Aut(\mathcal{C}) \leq AGL(1,p)$.
Furthermore $Aut(\mathcal{C})$ contains a normal $p$-Sylow group.
\item[(iii)]$Aut(\mathcal{C})=PSL(2,11)$; of degree $11$.
\item[(iv)]$Aut(\mathcal{C})=M_{11}$ or
$Aut(\mathcal{C})=M_{23}$ of degree $11$ or $23$, respectively.
\item[(v)] $S=PSL(d,r^{d^b})$ and $PGL (d,r^{d^b})\le  Aut(\mathcal{C}) \le P\Gamma L (d,r^{d^b})$ where
$d\in \mathbb{N}$, $d \geq 3$ is a prime number such that
$(d,r-1)=1$, and $p=(r^{d^{b+1}}-1)/(r^{d^b}-1)$.
\end{itemize}
\end{thm}
\pf A transitive group of prime degree is a primitive group~\cite[p.
195]{robinson}. As a consequence of a result of
Burnside~\cite[Theorem 2]{dobson}, a transitive group of prime
degree is either a subgroup of $AGL(1,p)$ or a doubly-transitive
group. Since the order of $AGL(1,p)$ is $p(p-1)$, the order of any
subgroup is $pm$ where $m|(p-1)$. By Sylow's Theorem,
$Aut(\mathcal{C})$ contains a unique $p$-Sylow group, so it is a
normal subgroup. By~\cite[Ex. 3.5.1]{dixon} $G$ is solvable. The
remaining cases follow from~Lemma~\ref{lem:sorley}.
A number theory argument~\cite[Lemma 3.1]{dixonzalesskii} gives that in case
(iv) if $p$ is prime, then $d$ must be a prime such that
$(d,r^a-1)=1$ and $a=d^b$.
The result then follows. \qed

%\pf A transitive group of prime degree is a primitive group~\cite[p.
%195]{robinson}. As a consequence of a result of
%Burnside~\cite[Theorem 2]{dobson}, a transitive group of prime
%degree is either a subgroup of $AGL(1,p)$ or a doubly-transitive
%group. In the first case $C_p \leq Per(C) \leq AGL(1,p)$, and if
%$p=2$ or $3$, $AGL(1,p)=S_p$. In this case, $C$ is elementary by
%Remark \ref{rem:iso2}, which is a contradiction. Since the order of
%$AGL(1,p)$ is $p(p-1)$, the order of any subgroup is $pm$ where
%$m|(p-1)$. By Sylow's Theorem, $Per(C)$ contains a unique $p$ Sylow
%group, so it is a normal subgroup. By~\cite[Ex. 3.5.1]{dixon} $G$ is
%solvable.
%Part (iv), we have from Lemma~\ref{lem:proj} that $Per(C)=P\Gamma
%L(d,t),$ $t=r^a$ for some $a \ge 1$ and $d\geq 3$. A number theory
%argument~\cite[Lemma 3.1]{dixonzalesskii} gives that if $p$ is
%prime, then $d$ must be a prime such that $(d,r^a-1)=1$ and $a=b^d$.
%The result then follows. \qed
%\begin{remark}
%It is conjectured that there are infinitely many primes of the form
%described in Theorem~\ref{lem:bur}(iii) (for example every Mersenne
%number has this form). See~\cite[p. 99]{dixon} for a more detailed
%discussion on these numbers. \qed
%3^9-1/3^3-1=757 premier
%\end{remark}
The following result is obtained by considering the automorphism
groups of cyclic objects of composite length.
\begin{thm}
\label{gr:p^r:primitif} Let $\mathcal{C}$ be a cyclic combinatorial
object on $n$ elements such that $n$ is a composite number.
Then $Aut(\mathcal{C})$ is either
\begin{enumerate}
\item[(i)] an imprimitive group
(in the case that $n=p^m$, $p$ prime, the orbit of the subgroup
generated by $T^{p^{m-1}}$ and its conjugate form a complete block system of $Aut(\mathcal{C})$);

 or
\item[(ii)] $Aut(\mathcal{C})$ is a doubly-transitive group such that
\[
PGL(d,r^a) \le Aut(\mathcal{C}) \le P\Gamma L(d,r^a),\text{ with }n =\frac{r^{ad}-1}{r^a-1}
\mbox{ and }a\ge 1.
\]
\end{enumerate}
 \end{thm}
\pf The group $Aut(\mathcal{C})$ contains a complete cycle and has
composite degree. Hence from a theorem of Burnside and
Schur~\cite[p. 65]{wielandt}, $Aut(\mathcal{C})$ is either
imprimitive or doubly-transitive.
If it is imprimitive and $n=p^m$, by \cite[Ch. XVI Theorem
VIII]{burnside2} $Aut(C)$ contains an intransitive normal subgroup
generated by $T^{p^{m-1}}$ and its conjugates. By~\cite[Proposition
7.1]{wielandt} the orbit of such a subgroup forms a complete block
system of $Aut(\mathcal{C})$.

In the doubly-transitive case, we have from Lemma~\ref{lem:sorley}
that the only cases when the socle can be abelian are $N=C_p$ and
$N=C_2 \times C_2$. In these cases, $Aut(\mathcal{C})$ must be equal
to $AGL(1,p)$ or $S_4$, which is impossible. Since the socle is not
abelian and the degree is not prime, this leads to the only solution
given by row six of Table 1.
% in Lemma~\ref{lem:sorley}.
\qed
\begin{cor}(\cite[Corollary 8.6]{joy})
\label{cor:joy}
For any circulant graph $\mathcal{C}$ on $n$ elements, one of the following holds:
\begin{enumerate}
\item $Aut(\mathcal{C})=S_n$;
\item $Aut(\mathcal{C})$ is imprimitive, and the orbit of the subgroup generated by
$T^{p^m-1}$ and its conjugate form a complete block system of
$Aut(\mathcal{C})$; or
\item $n$ is prime and $Aut(\mathcal{C}) < AGL(1,p)$.
\end{enumerate}
\end{cor}
\pf
If an automorphism group acting on a graph is doubly transitive, then it takes any ordered pair
of vertices to another ordered pair of vertices.
Hence the graph is either complete or empty, and (in either case) the automorphism group is $S_n$.
The result then follows from Theorems~\ref{lem:bur} and~\ref{gr:p^r:primitif}.
\qed

\section{Automorphism Groups of Cyclic Codes}
A linear code $C$ over $\mathbb{F}_q$ is cyclic if $T \in
Aut(\mathcal{C})$, where $T=(1,2,\ldots,n)$ is a complete cycle of
length $n$.  In the case of cyclic codes we have the following
results concerning the group $Aut(\mathcal{C})$.
We begin with the following useful remark.
\begin{rem}
\label{rem:iso2} The zero code, the entire space, and the repetition
code and its dual are called elementary codes. The permutation group
of these codes is $S_n$~\cite[p. 1410]{hand}. Furthermore, it was
proven in \cite[p. 1410]{hand} that there is no cyclic code with
permutation group equal to $Alt(n)$. \qed
\end{rem}
The following lemma can be proved using arguments similar to those for the binary case~\cite[Theorem E Part 3]{rolf}.
\begin{lem}
\label{lem:proj} Let $\mathcal{C}$ be a non-elementary cyclic code
of length $n=\frac{t^d-1}{t-1}$ over a finite field $\mathbb{F}_q$,
where $q=r^{\alpha}$ and $t$ is a prime power.
If $Aut(\mathcal{C})$ satisfies
\[
PGL(d,t) \leq Aut(\mathcal{C}) \leq P\Gamma L(d,t),
\]
then $t=r^a$ for some $a \ge 1$, $d\geq 3$, and $Aut(\mathcal{C})=P\Gamma L(d,t)$.
\end{lem}
\pf Assume $d=2$. As the group $PGL(2,t)$ acts 3-transitively on the
$1$-dimensional projective space $\mathbb{P}^1(\mathbb{F}_t)$, we
deduce from~\cite[Table 1 and Lemma 2]{mortimer} that the
underlying code is elementary, which is a contradiction. Hence
$d\geq 3,$ and from~\cite[Table 1 and Lemma 2]{mortimer}, it must be
that since $C$ is non-elementary, $t$ must be equal to $r^a$. Now
let $V$ denote the permutation module over $ \mathbb{F}_{p}$
associated with the natural action of $PGL(d,t)$ on the $(d-1)$
dimensional projective space $\mathbb{P}^{d-1}( \mathbb{F}_t)$. Let
$U_1$ be a $PGL(d,t)$-submodule of $V$. Hence $U_1$ is $P\Gamma
L(d,t)$-invariant. This is because, if $\sigma$ is a generator of
the cyclic group $P\Gamma L(d,t)/PGL(d,t)
 \simeq Gal( \mathbb{F}_t/ \mathbb{F}_{p})$,
then $U_2=U_1^{\sigma},$ regarded as a $PGL(d,t)$-module, is simply
a twist of $U_1$. Let $\overline{\mathbb{F}}_{r}$ be the algebraic
closure of $\mathbb{F}_{r}$. Then the composition factors of the
$\overline{\mathbb{F}}_{p}PGL(d,t)$-modules
$\overline{U}_1=\overline{\mathbb{F}}_{r}\otimes U_1$ and
$\overline{U}_2=\overline{\mathbb{F}}_{r}\otimes U_2$ are the same.
The submodules of the $\overline{\mathbb{F}}_{r}PGL(d,t)$-module
$\overline{V}=\overline{\mathbb{F}}_{r}\otimes V$ are uniquely
determined by their composition factors~\cite{bardoe}. Then we have
$\overline{U}_1=\overline{U}_2$, which implies that $U_1=U_2$, and
therefore $Aut(\mathcal{C})=P\Gamma L(d,t)$.
\qed
\begin{thm}
\label{lem:bur2} Let $\mathcal{C}$ be a non-elementary cyclic code
of length $n$ over $\mathbb{F}_{q}$, where $q=r^{\alpha}$, $\alpha
\ge 1$. Then if $n=p$ is a prime number we have that $Aut(\mathcal{C})$ is a
primitive group, and one of the following holds:
\begin{itemize}
\item[(i)] $Aut(\mathcal{C})$ is a solvable group of order $pm$ with $m$ a divisor of $p-1$ and $C_p \leq Aut(\mathcal{C}) \leq AGL(1,p)$, with $p \geq 5$.
Furthermore $Aut(\mathcal{C})$ contains a normal $p$-Sylow group.
\item[(ii)] If $p=q$, then $Aut(\mathcal{C})=AGL(1,p)$.
\item[(iii)] $Aut(\mathcal{C})=PSL(2,11)$ and $q$ is a power of 3.
$C$ is either an $[11,6]$ or $[11,5]$ code that is equivalent to the $[11,6,5]$ ternary Golay code or its dual, respectively.
\item[(iv)] $Aut(\mathcal{C})=M_{23}$ and $q$ is a power of 2.
$C$ is either a $[23,12]$ or $[23,11]$ code that is equivalent to the $[23,12,7]$ binary Golay code or
its dual, respectively.
\item[(v)] $Aut(C)=P\Gamma L (d,r^{d^b})$ where $b\in \mathbb{N}$, $d \geq 3$ is a prime number such that $(d,r^{d^b}-1)=1$, and $p=(r^{d^{b+1}}-1)/(r^{d^b}-1)$.
\end{itemize}
\noindent
If $n$ is a composite number then $Aut(\mathcal{C})$ is either
\begin{itemize}
\item[(vi)] an imprimitive group
(in the case that $n=p^m$, $p$ a prime, the orbit of the subgroup
generated by $T^{p^{m-1}}$ and its conjugate form a complete block
system of $Aut(\mathcal{C})$);

\noindent
or
\item[(vii)] $Aut(\mathcal{C})$ is a doubly-transitive group
equal to $P\Gamma L(d,r^a),\text{ with }n =\frac{r^{ad}-1}{r^a-1} \mbox{ and }
a\ge 1$.
\end{itemize}
\end{thm}
\pf From Theorem~\ref{lem:bur} we have that $Aut(\mathcal{C})$ is
either a subgroup of $AGL(1,p)$ which is solvable of order $pm$, $m$
a divisor of $p-1$, or a doubly transitive group. For the first case
if $p=2$ or $3$, then $AGL(1,p)=S_p$, and $\mathcal{C}$ is elementary by Remark~\ref{rem:iso2}.
For part (ii), if $q=p$,
Roth and Seroussi~\cite{roth} proved that any cyclic code of prime
length $p$ over $\mathbb{F}_p$ must be an MDS code equivalent to an extended Reed--Solomon code.
Berger \cite{berger91} proved that the permutation group of such codes is the affine group $AGL(1,p)$.
For parts (iii) and (iv), as $\mathcal{C}$ is non-elementary of prime
length $p$, by Lemma~\ref{lem:sorley}, Remark~\ref{rem:iso2} and
Lemma~\ref{lem:proj}, we have that $Aut(\mathcal{C})$ is one of
$M_{11}$ with $p=11$, $PSL(2,11)$ with $p=11$, $M_{23}$ with
$p=23$, or $ P\Gamma L(d,t)$ of degree $p=(t^d-1)/(t-1)$ and $t$ a prime power.
If $Aut(\mathcal{C})=M_{11}$, from~\cite[Table 1, Lemma 2]{mortimer} $C$ must be elementary, which is a contradiction.
If $Aut(C)=PSL(2,11)$, from~\cite[Table 1, Lemma 2 and (J)]{mortimer} $q$ must be a power of 3, and there is a unique
non-elementary code over $\mathbb{F}_q$ contained in the dual of the repetition code.
The $[11,5,6]$ dual of the ternary Golay code is contained in the repetition code and has permutation group
$PSL(2,11)$; its dual, an $[11,6,5]$ code, intersects the dual of
the repetition code in this $[11,5,6]$ code and also has permutation group $PSL(2,11)$.
Part (ii) then follows.
Part (iii) is obtained in an analogous way from~\cite[Table 1, Lemma 2 and (I)]{mortimer}.
\qed

Now we give an algorithm to find the automorphism group of specific
cyclic codes. Let $\mathcal{C}$ be a non elementary cyclic code over
$\F_r$ of length $p$, different from the binary and ternary
Golay codes. Further, assume that there are no integers $b$ and $d$
such that $p=(r^{d^{b+1}}-1)/(r^{d^b}-1)$.
Then from~Theorem~\ref{lem:bur} and Remark~\ref{rem:iso}, we have that $\mathcal{C} \le AG(p)$.
Let $a \in \Z_p^*$ and $\mu_a$ be the associated multiplier.
Hence if $\mu_a(\mathcal{C})=\mathcal{C}$ then $\mu_a \in Aut(\mathcal{C})$.
From Remark~\ref{rem:multi}, for all $b\in \Z_p$ we also have that $\tau_{a,b} \in Aut(\mathcal{C})$.
This suggest the following algorithm to find $Aut(\mathcal{C})$.
To summarize, it is assumed that $\mathcal{C}$ is not elementary, $p\neq 11$ and $p \neq 23$,
and there are no integers $b$ and $d$ such that $p=(r^{d^{b+1}}-1)/(r^{d^b}-1)$.
\hfill \\

\noindent
\textbf{Algorithm A}:
\begin{enumerate}
 \item Find $A=\{a \in \Z_p^* \text{ such that }\mu_a (\mathcal{C})=\mathcal{C}\}$.
\item If $A =\Z_p^*$, then $Aut(\mathcal{C})=AG(p)$.
\item Otherwise, $Aut(\mathcal{C})=\{\tau_{a,b}, a \in A, b \in \Z_p\}$.
\end{enumerate}
%\begin{ex}
%The group $P\Gamma L(3,3)$ is the automorphism group of the ternary
%narrow-sense BCH code of length 13 and designed distance 4.
%\end{ex}
In Table 2, we give examples of permutation groups of BCH codes of
length $p$ over $\mathbb{F}_q$. $Aut(C)$ (respectively $Aut(C_2)$
and $Aut(C_3)$), denote the permutation groups of narrow sense
($b=1$) BCH codes with designed distance $\delta$ (respectively BCH
codes with designed distance $\delta$ and $b=2$ and $b=3$).
\begin{table}
\begin{center}
\begin{tabular}{|c|c|c|c|c|c|}
\hline
 $q$&$p$& $\delta$& $Aut(C)$&$Aut(C_2)$& $Aut(C_3)$\\
\hline
2&17&2&$C_8 \ltimes C_{17}$&$S_{17}$&$S_{17}$\\
2&23&3&$M_{23}$& $M_{23}$ & $M_{23}$\\
2&41&2&$C_{20} \ltimes C_{41}$&$C_{20} \ltimes C_{41}$&$C_{20} \ltimes C_{41}$ \\
2&41&3&$C_{20} \ltimes C_{41}$&$S_{41}$&$S_{41}$\\
2&43&5&$C_{14} \ltimes C_{43}$&$C_{14} \ltimes C_{43}$&$C_{14} \ltimes C_{43}$\\
2&43&7&$C_{14} \ltimes C_{43}$&$S_{43}$&$S_{43}$\\
3&13&2&$C_3 \ltimes C_{13}$&$C_3 \ltimes C_{13}$ &$C_3 \ltimes C_{13}$\\
3&13&4&$P\Gamma L(3,3)$& $C_3 \ltimes C_{13}$&$C_3 \ltimes C_{13}$\\
3&13&5&$C_3 \ltimes C_{13}$&$C_3 \ltimes C_{13}$&$C_3 \ltimes C_{13}$\\
3&23&3&$C_{11} \ltimes C_{23}$ &$C_{11} \ltimes C_{23}$&$C_{11} \ltimes C_{23}$\\
3&41&5&$C_{8} \ltimes C_{41}$&$C_{8} \ltimes C_{41}$&$C_{8} \ltimes C_{41}$\\
4&43&9&$C_{7} \ltimes C_{43}$&$S_{43}$&$S_{43}$\\
5&11&5&$C_{5} \ltimes C_{11}$&$C_{5} \ltimes C_{11}$&$C_{5} \ltimes C_{11}$\\
11&5&3&$C_{5}$&$C_{2} \ltimes C_{5}$& $C_{5}$\\
\hline
\end{tabular}
\end{center}
\caption{Permutation Groups of some BCH Codes of Length $p$}
\label{table2}
\end{table}
%\begin{rem}
%If $\mathcal{C}$ is a circulant graph on $p$ vertices with $p$ a
%prime number, hence from Theorem~\ref{lem:bur} and
%Corollary~\ref{cor:joy} $Aut(\mathcal{C})$ is a subgroup of $AG(p)$
%or $S_p$. Hence we can determine $Aut(\mathcal{C})$ by an adaptable
%version of Algorithm A. This has also been given by Morris~\cite[p.
%5]{joy}
%\end{rem}

\subsection{The Automorphism Groups of Cyclic Codes of Length $p^m$}

In the previous section, the automorphism groups of cyclic codes were determined.
In this section, we provide additional results on these cyclic
combinatorial objects in the case $n=p^m$, where $p$ is an odd prime and $m \ge 1$.
%It is isomorphic with the wreath
%product $\mathbb{Z}_p \wr \ldots \wr \mathbb{Z}_p$
\begin{lem}
\label{lem:ord} Let $q$ be a prime power, $p$ an odd prime, and $z$
the largest integer such that $p^z | (q^t-1)$, with $t$ the order of $q$ modulo $p$.
If $z=1$ we have
\[
\ord_{p^m}(q)=p^{m-1}t.
\]
\end{lem}
\pf Let $t$ be the order of $q$ modulo $p$, and $u=q^t \equiv 1 \bmod
p$. Assume that $z=1$, or equivalently $u \neq 1 \bmod p^2$. It is
well known from elementary number theory~\cite[p. 87]{demazure} that
$u \bmod p^{m}$ is an element of order $p^{m-1}$ in the group
$\left(\mathbb{Z}_{p^{m}}\right) ^{\ast }$ if and only if $u \neq 1
\bmod p^2$. Hence $\ord_{p^m}(q)=p^{m-1}t$. \qed

\noindent Note that according to Brillhart et al.~\cite{brillhart}, it is unusual to have $z >1$.
\begin{prop}
\label{lem:groupe}
Let $\mathcal{C}$ be a cyclic object on $p^m$ elements with $ m >1$.
Hence a $p$-Sylow subgroup of $Aut(\mathcal{C})$ has order $p^s$ such that
\begin{equation}
m \leq s \leq p^{m-1}+p^{m-2}+ \cdots +1.
\end{equation}
We consider $\mathcal{C}$ to be a cyclic code of length $p^m$ over $\F_q$ with $q=r^{\alpha}$ a prime power and $(q,p)=1$.
Let $\mu_q$ be the multiplier defined by $\mu_q(i)=iq \bmod p^m$.
Then the group $Aut(\mathcal{C})$ contains the subgroup $K=<T,\mu_{q}>$ of order $p^m\ord_{p^m}(q)$.
Let $p^l$, $l \geq m$, be the $p-$part of the order of $K$.
Then a $p$-Sylow subgroup $P$ of $Aut(\mathcal{C})$ has order $p^s$ such that
\[
l \le s \le  p^{m-1}+ p^{m-2}+ \cdots 1.
\]
If $z=1$, then $s \geq 2m-1$.
If $s=2m-1$, then $P$ is a transitive group of $K$.
\end{prop}
\pf From the definition of a cyclic code, we have that $T\in Aut(\mathcal{C})$.
It is obvious that each cyclotomic class modulo $n$ over $\mathbb{F}_q$ is invariant under the permutation $\mu_q$.
This can be deduced from the fact that the polynomial $f(x)\in \mathbb{F}_q[x]$ satisfies $f(x^q)=f(x)^q$.
Thus $\mu_q \in Aut(\mathcal{C})$.
The order of $\mu_q$ is equal to $|Cl(1)|= \ord_{p^m} (q)$, and hence $K=<T, M_q>$ is a subgroup of
$Aut(\mathcal{C})$ of order ${p^m} \ord_{p^m}(q)$.
Then the order of $K$ has $p-$part $p^l$ with $l\le m$.
Let $P$ be a $p$-Sylow subgroup of $Aut(\mathcal{C})$ which contains $T$ (this can always
be assumed since any $p$ group is contained in a $p$-Sylow subgroup).
Then $P$ is a $p$ group of $S_{p^m}$.
From Sylow's Theorem, $P$ is contained in a $p$-Sylow subgroup of $S_{p^m}$.
It is well known that a $p$-Sylow subgroup of $S_{p^m}$ has order
$p^{p^{m-1}+p^{m-2}+ \cdots +1}$~\cite[Kalu\v{z}nin's Theorem]{robinson}.
Since $P$ also contains the subgroup of $K$ of order $p^l$, then $l\leq s \leq p^{m-1}+p^{m-2}+ \cdots +1$.
If $z=1$, then by Lemma~\ref{lem:ord} the order of the group $K$ is $\ord_p(q)p^{2m-1}$.
This gives that $p^{2m-1}$ divides $|Aut(\mathcal{C})|$, so $Aut(\mathcal{C})$ contains a $p$ subgroup
of order at least $p^{2m-1}$.
If $s=2m-1$, we can assume that $P \leq K$ because we have $T \in K$ and $K \leq Aut(\mathcal{C})$.
Thus $P$ is a transitive subgroup of $K$. \qed
\begin{thm}
\label{th:Zsigmondy} Let $\mathcal{C}$ be a non elementary cyclic
code of length $p^m$ over $\mathbb{F}_{r^{\alpha}}$ with $m \geq 1$.
Then the following holds:
\begin{itemize}
\item[(i)] If $p \nmid \alpha$ and $p \nmid (d,r^a-1)$, then
$Aut(\mathcal{C})=P\Gamma L(d,r^a)$, $a\ge 1$, $d \ge 3$, if and
only if the $p$-Sylow subgroup of $Aut(C)$ is of order $p^m$.
\item[(ii)] If $p \geq 5$, $\alpha=1$ and $r=p$, $m>1$, then
$Aut(\mathcal{C})$ is an imprimitive group which admits a complete
system formed by the orbit of the subgroup generated by
$T^{p^{m-1}}$ and its conjugate.
 It also contains a transitive normal $p$-Sylow subgroup of order $p^s$ with $m<s \le
p^{m-1}+p^{m-2}+\ldots+1$.
\item[(iii)] If $z=1$, $p\nmid \alpha$ and $p \nmid (d,r^a-1)$, then
$Aut(\mathcal{C})$ is an imprimitive group which contains a
transitive normal $p$-Sylow subgroup of order $p^s$, with $2m -1 \le
s \le p^{m-1}+p^{m-2}+\ldots +1$. Furthermore, $Aut(\mathcal{C})$
admits a complete block system formed by the orbit of the subgroup
generated by $T^{p^{m-1}}$ and its conjugate.
\end{itemize}
\end{thm}
\pf For part (i), we know that the socle of $P\Gamma L(d,r^a)$ is
the group $PSL(d,r^a)$ of order
$\frac{r^{ad(d-1)/2}}{(d,r^a-1)}\prod_{i=2}^{d}(r^{ai}-1)$. From a
lemma of Zsigmondy~\cite[Ch. IX, Theorem 8.3]{huppert}, except for
the cases $d=2$, $r^a=2^b-1$ and $d=6, r^a=2$, there exists a prime
$q_0$ such that $q_0$ divides $r^{ad}-1$, but does not divide
$r^{ai}-1$, for $1\le i < d$. From Lemma~\ref{lem:proj}, we cannot
have $d=2$. The case $d=6$ and $r^a=2$ does not give a prime power.
Hence if $n=p^m=\frac{r^{ad}-1}{r^{a}-1}$, there is a $q_0$ which
divides $(r^{ad}-1)=(r^a-1)p^m$. Since $q_0$ does not divide
$r^a-1$, $q_0$ must divide $p^m$, and hence $q_0=p$ and $p^m$ is
the $p-$part of the order of $PSL(d,r^a)$. Also, since $p \nmid
r^a-1$, we have that $p \nmid (d,r^a-1)$. Hence if $(\alpha,p)=1$,
$p^m$ is also in the $p-$part of the order of $P\Gamma L(d,r^a)$,
and the result follows.

Conversely, if $Aut(\mathcal{C})$ contains a $p$-Sylow group $P$ of order
$p^m$, we can assume that $T\in P$, which gives the equality
$P=\langle T \rangle $. Assume that in this case $Aut(\mathcal{C})$
is imprimitive. Then by \cite[Theorem 33]{dobson}, $P$ is normal.
$P$ is then the minimal normal subgroup which is transitive and
abelian. From~\cite[p. 17]{wielandt} $Aut(\mathcal{C})$ is
primitive, which is impossible. Thus if $P=\langle T \rangle$, the
group $Aut(\mathcal{C})$ is equal to $P\Gamma L(d,r^a)$, which is
possible only if $[P\Gamma L(d,r^a):PSL(d,r^a)]$ is prime to $p$,
i.e., $(p,\alpha)=1$ and $p\nmid (d,r^a-1)$.

For part (ii), from Theorem~\ref{gr:p^r:primitif} if
$Aut(\mathcal{C})$ is primitive, then it is doubly-transitive and
equal to $P\Gamma L(d,r^a)$ with $n =\frac{r^{ad}-1}{r^a-1}, d\geq
3$ and $a \ge 1$. From~\cite[Lemma 22]{dobson}, if
$Aut(\mathcal{C})$ is doubly-transitive with a non abelian socle, then
$Soc(Aut(\mathcal{C}))=Alt(p^m)$. Hence from Remark~\ref{rem:iso2}
the code is elementary. Since $Aut(\mathcal{C})$ is imprimitive,
from part (i) the order of the $p$-Sylow group is $p^s$ with $s>m$.
The second inequality then follows from~Proposition~\ref{lem:groupe}.

For part (iii), if $z=1$ then from Proposition~\ref{lem:groupe},
the order of a $p$-Sylow subgroup of $Aut(\mathcal{C})$ is at least $p^{2m-1}$.
If $Aut(\mathcal{C})$ is doubly transitive, by
Theorem~\ref{gr:p^r:primitif} it is equal to $P\Gamma L(d,r^a)$ with $d \geq 3$.
By assuming $p\nmid \alpha$ and $p\nmid (d,r^a-1)$, we obtain from part (i) that
a $p$-Sylow group of $Aut(C)$ has order $p^m$, which is impossible.
Hence $Aut(\mathcal{C})$ is an imprimitive group.
From~\cite[Theorem 33]{dobson}, $Aut(\mathcal{C})$
contains a transitive normal $p$-Sylow subgroup.
The result then follows.
\qed
\begin{ex}
The narrow sense BCH code of length 25 over $\mathbb{F}_3$ with
designed distance 3 has an automorphism group which is the
imprimitive group $S_5 \wr S_5$. The narrow sense BCH code of length
9 over $\mathbb{F}_5$ with designed distance 2 has an automorphism
group which is the imprimitive group $S_3 \wr S_3$. The binary
$[7,4,3]$ Hamming code has automorphism group $P\Gamma L(3,2)$,
which contains a $7$-Sylow subgroup of order 7.
\end{ex}
%\begin{remark}
%It is a well known fact that the automorphism group of the binary
%Hamming code of length $2^d-1$ is $PSL(d,2) =P\Gamma L(d,2)$. The
%Theorem~\ref{gr:p^r:primitif} (ii) generalizes this fact.
%\end{remark}

\section{Equivalence of Cyclic Combinatorial Objects on $p^m$ Elements}

Let $\mathcal{C}$ be a cyclic object of length
$p^m$ where $p$ is an odd prime, $m>1$ and $P$ is a $p$-Sylow subgroup of $Aut(\mathcal{C})$.
The following subset of $S_{p^r}$ was introduced by Brand~\cite{brand}
\[
H(P)=\{ \sigma \in S_{p^m} |  \sigma ^{-1} T \sigma \in P \}.
\]
The set $H(P)$ is well defined since $<T>$ is a subgroup of
$Aut(\mathcal{C})$ of order $p^m$, hence it is a $p$-group of $Aut(\mathcal{C})$.
From Sylow's Theorem, there exists a $p$-Sylow subgroup $P$ of $Aut(\mathcal{C})$ such that $<T> \; \leq P$.
Furthermore, in some cases the set $H(P)$ is a group.

\begin{lem}(\cite[Lemma 3.1]{brand})
\label{lem:brand} Let $\mathcal{C}$ and $\mathcal{C}'$ be cyclic
objects on $p^m$ elements. Let $P$ be a $p$-Sylow subgroup of
$Aut(\mathcal{C})$ which contains $T$. Then $\mathcal{C}$ and
$\mathcal{C}'$ are equivalent if and only if $\mathcal{C}$ and
$\mathcal{C}'$ are equivalent by an element of $H(P)$.
\end{lem}

Let $p$ be an odd prime.
For $n<p$, we define the following subsets of $S_{p^m}$:
\[
\begin{array}{l}
Q^n=\{f: \mathbb{Z}_{p^m} \rightarrow \mathbb{Z}_{p^m} | {\displaystyle f(x)=\sum_{i=0}^{n}a_ix^i}, a_i \in \mathbb{Z}_{p^m}
\mbox{ for each }i, (p,a_1)=1,\\
\hspace*{0.5in} \mbox{ and }p^{m-1}\mbox{ divides }a_i\mbox{ for }i=2,3,\ldots, n\}.
\end{array}
\]
\[
Q_1^n=\{f \in Q^n | f(x)=\sum_{i=0}^{n}a_ix^i, \text{ with }a_1 \equiv 1 \bmod p^{m-1} \}.
\]
The sets $Q^n$ and $Q_1^n$ are subgroups of $S_{p^m}$~\cite[Lemma 2.1]{brand}. Note that $Q^1=AG(p^m)$.

\begin{lem}
\label{lem:odd}
Let $\mathcal{C}$ be a cyclic object on $p^m$ elements, where $p$ is odd and $m>1$.
Let $P$ be a Sylow subgroup of $Aut(\mathcal{C})$ which contains $T$.
If $1\leq n<p$, then
\begin{itemize}
\item[(i)] $|Q^n|=(p-1)p^{2m+n-2}$ and $|Q_1^n|=p^{m+n}$.
\item[(ii)] $AG(p^m)=N_{S_{p^m}}(<T>) \subset H(P)$ .
\item[(iii)] $Q^{n+1}=H(Q_1^n)$.
\item[(iv)] $N_{S_{p^m}}(Q_1^n)=Q^{n+1}$.
\end{itemize}
\end{lem}
\pf For part (i), from~\cite[Lemma 3.2]{brand} we have the map
$(a_0,\ldots,a_n) \longrightarrow f$ where $f(x)= \sum_{i=0}^n a_i x^i$ is injective if $n<p-1$.
Thus in $Q^n$, the coefficients of
$a_0$ can take $p^m$ different values, and $a_1$ can take
$p^{m-1}(p-1)$ values. For $2\leq i \leq n$, $a_i$ can take $p$
values. From these results we have $|Q^n|=p^{2m+n-2}(p-1)$. For
$Q_1^n$, the coefficients of $a_0$ can take $p^m$ different values,
and $a_i$ for $1\leq i\leq n$ can take $p$ values, hence $|Q_1^n|=p^{m+n}$.

Now we prove that $AG(p^m)=N_{S_{p^m}}(<T>)$. Let $\sigma$ be an
element of $N_{S_{p^m}}(<T>)$. Then there is a $j \in \mathbb{Z}_n
\setminus \{0\}$ such that $\sigma T \sigma^{-1}=T^j$, or
equivalently $\sigma T=T^j \sigma$. Hence $\sigma T(0)=\sigma (1)=
T^j \sigma(0)=\sigma(0)+j$ and $\sigma T(1)=\sigma
(1)+j=\sigma(0)+2j$, so that $\sigma (k)=\sigma(0)+kj$ for any $k \in
\mathbb{Z}_n$. Then $(j,n)=1$ follows from the fact that the order
of $T$ equals the order of $T^j$. The last inclusion is obvious.

Part (iii) follows from~\cite[Lemma 3.7]{brand}.

For the proof of part (iv),
we begin with the $\leq$ condition.
Let $h\in N_{p^m}(Q_1^n)$ and $g=h^{-1}T h$.
As $T\in Q_1^n$, it must be that $g\in Q_1^n$. Since the order of $g$
is equal to the order of $T$ which is $p^m$, from~\cite[Lemma
3.6]{brand} there exists $f\in Q^{n+1}$ such that $f^{-1}gf=T$. Thus
$f^{-1}h^{-1}Thf=T$. The only elements of $S_{p^m}$ which commute
with $T$ (a complete cycle of length $p^m$), are the powers of $T$.
Thus $hf=T^j$ for some $j$.
Since $Q^{n+1}$ is a subgroup of $S_{p^m}$ and $\langle T \rangle \leq Q^{n+1}$,
then $ h \in Q^{n+1}$, and hence $N_{p^m}(Q_1^n)\leq Q^{n+1}.$

Now consider the $\geq$ condition.
Let $h \in Q_1^n$, where $h(x)=\sum_{i=0}^{n}h_ix^i$
with $h_1 \equiv 1 \bmod p^{m-1}$ and $p^{m-1} |h_i$ for $2 \leq i \leq n$.
Let $g \in Q^{n+1}$ where $g(x)=\sum_{i=0}^{n+1}g_ix^i$ with $p \nmid  g_1$ and
$p^{m-1} |g_i$ for $2 \leq i \leq n$.
We have
\[
hg(x)=\sum_{i=0}^{n}h_i\left(\sum_{j=0}^{n+1}g_jx^j\right)^i=h_0+h_1 \sum_{i=0}^{n+1}g_jx^j+\sum_{i=2}^{n}h_i\left(\sum_{j=0}^{n+1}g_jx^j\right)^i.
\]
Since $p^{m-1} |h_i,$ for $i \geq 2$ and $p^{m-1} | g_j$ for $j \geq
2$, any terms in $\sum_{i=2}^{n}h_i\left(\sum_{j=0}^{n+1}g_jx^j\right)^i$
involving $g_j$ for $j \geq 2$ vanish modulo $p^m$, so that
\[
hg(x)=h_0+h_1\sum_{j=0}^{n+1}g_jx^j+\sum_{i=2}^{n}h_i\left(g_{0}+g_1x\right)^i.
\]
By \cite[Lemma 2.1]{brand}
\begin{equation}
\label{eq:inverse} g^{-1}(x)=\sum_{i=1}^{n+1}b_ix^i, \text{ with
}b_1 =g_1^{-1} \text{ and }b_i=-g_ig_1^{-(i+1)}\text{ for } 2 \leq j
\leq n+1.
\end{equation}

We now determine $g^{-1}hg$ in order to prove that it is in $Q_1^n$.
This is given by
\[
g^{-1}hg(x)=\sum_{k=1}^{n+1}b_k\left(h_0+h_1\sum_{j=0}^{n+1}g_jx^j+
\sum_{i=2}^{n}h_i(g_0+g_1x)^i-g_0\right)^k
\]
\[
=b_1\left(h_0 +h_1 \sum_{j=0}^{n+1}g_jx^j+\sum_{i=2}^{n}h_i\left(g_0+g_1x\right)^i-g_0\right)
\]
\[
\hspace*{0.5in} +\sum_{k=2}^{n+1}b_k{\left(h_0+h_1\sum_{j=0}^{n+1}g_jx^j+\sum_{i=2}^{n}
h_i {(g_0+g_1 x)}^i-g_0\right)}^k.
\]
As $p^{m-1}|g_j$ for $j \geq 2$, hence $p^{m-1}|b_k$ for $k \geq 2$.
Furthermore, we have $p^{m-1}|h_i$ for $i \geq 2$, and thus
\[
g^{-1}hg(x)=b_1\left(h_0+h_1\sum_{j=0}^{n+1}g_jx^j+\sum_{j=0}^{n+1}h_i(g_0+g_1x)^i-g_0\right)+
\sum_{k=2}^{n+1}b_k\left(h_0+h_1\left(g_0+g_1x\right)-g_0\right)^k.
\]
Let $g^{-1}hg(x)=\sum_{m=0}^{n+1}c_mx^m$, and note that
$c_{n+1}=b_1h_1g_{n+1}+b_{n+1}(h_1g_1){n+1}$.
Then replacing the $b_i$ with their values from (\ref{eq:inverse}), we obtain
$$c_{n+1}=g_1^{-1}h_1g_{n+1}-g_{n+1}g_1^{-(n+2)}h_1^{n+1}g_1^{n+1}=g_1^{-1}h_1(g_{n+1}-g_{n+1}h_1^n).$$
As $h_1\equiv 1 \bmod p^{m-1}$, we have that $h_1^n \equiv 1 \bmod p^{m-1}$.
In addition, as $p^{m-1}|g_{n+1}$, it must be that $g_{n+1}h_1^n \equiv g_{n+1} \bmod p^m$.
Therefore, $c_{n+1}=0$, and $p^{m-1} |c_i$ for $2\leq i\leq n$.
Then we only need to show that $c_1 \equiv 1 \bmod p^{m-1}$.
As $g_{j}\equiv 0 \bmod p^{m-1}$ for $j \geq 2$, $h_i\equiv 0 \bmod p^{m-1}$ for $i \geq 2$,
and $b_k \equiv 0 \bmod p^{m-1}$ for $k \geq 2$, then $c_1\equiv b_1h_1g_1 \bmod p^{m-1}$.
Finally, since $b_1=g_1^{-1}$, we have that $c_1\equiv h_1 \equiv 1 \bmod p^{m-1}$.
\qed

\begin{lem}
\label{lem:2.6} Let $1\leq n <p-1$. If $P$ is a $p$ group of
$S_{p^m}$ with $Q_1^n \lneq P \leq Q^{n+1}$, then $P=Q_1^{n+1}$.
\end{lem}
\pf By part (ii) of Lemma~\ref{lem:odd}, we have $Q_1^n \lhd Q^{n+1}$.
Hence we can consider $\overline{Q}=Q^{n+1}/Q_1^n$, which has
order $p^{m-1}(p-1)$ by Lemma~\ref{lem:odd}. Let $N$ be the number
of $p$-Sylow subgroups of $\overline{Q}$. Then by Sylow's Theorem,
$N \equiv 1 \bmod p$ and $N$ divides $p^{m-1}(p-1)$. Hence $N=1$, so
there exists a unique $p$-Sylow subgroup $\overline{P'}$ of
$\overline{Q}$ which is normal. From the condition on $P$ above, the
image $\overline{P}$ of $P$ in $\overline{Q}$ is also a $p$-Sylow
subgroup of $\overline{Q}$. Since there is a unique $p$-Sylow
subgroup $\overline{P'}=\overline{P}$, by Lemma~\ref{lem:odd} the
image $\overline{Q}_1^{n+1}$ of $Q_1^{n+1}$ in $\overline{Q}$ is a
$p$-Sylow subgroup of $\overline{Q}$. Hence
$\overline{Q}_1^{n+1}=\overline{P}=\overline{P'}$. As $Q_1^{n}\lneq
P$ and $Q_1^{n} \leq Q_1^{n+1}$, the result follows. \qed

Now we prove that the group $Q_1^1$ is a special subgroup of
$S_{p^m}$.
\begin{thm}
\label{lem:unique} The group $Q_1^1$ is a normal subgroup of $Q^1$
and is the unique subgroup of $S_{p^m}$ of order $p^{m+1}$ which
contains $T$.
\end{thm}
\pf It is obvious that $T \in Q_1^1$, and from Lemma~\ref{lem:odd},
$|Q_1^1|=p^{m+1}$. Consider now an element $g$ of $Q^1$,
$g(x)=b_0+b_1x$, with $b_0,b_1 \in \mathbb{Z}_{p^m}$ and $(b_1,p)=1$.
It is not difficult to determine that the inverse of $g$ in $Q^1$ is
given by $g^{-1}(x)=-b_1^{-1}b_0+b_1^{-1}x$. Consider $f\in Q_1^1$,
so that $f(x)=a_0+a_1x$ with $a_0,a_1 \in \mathbb{Z}_{p^m}$,
$(a_1,p)=1$ and $a_1 \equiv 1 \bmod p^m$. We then have
$g^{-1}fg(x)=g^{-1}(a_0+a_1(b_0+b_1x))=(-b_0+a_0+a_1b_0)b_1^{-1}+a_1x$.
This proves that $ g^{-1}fg(x) \in Q_1^1$. Hence $Q_1^1$ is normal
in $Q^1$. Now let $S$ be a subgroup of $Q^1$ of order $p^{m+1}$
which contains $T$. Thus $<T>$ has index $p$ in $S$, and so $<T>$
is maximal in $S$. Furthermore, $<T> \lhd S$, because any subgroup
of a $p$-group of index $p$ must be normal.
%~\cite[p. 7 Exercise1.3.4]{dixon}.
Therefore, we have $S=N_S(T) \leq N_{S_{p^m}} (T)$, and by
Lemma~\ref{lem:odd}, $S \leq N_{S_{p^m}} (T)=AG(p^m)=Q^1$. Thus,
such an $S$ must be a subgroup of $Q^1$. It is clear that $Q_1^1$ is
not abelian, and $S$ cannot be abelian since it is a transitive
group. If this were the case it would have to be a regular
group~\cite[Theorem 1.6.3]{robinson}, and thus $|S|=p^m$, which is
impossible. Furthermore, the $p$ groups which contain a cyclic
maximal subgroup are known~\cite[Theorem 5.3.4]{robinson}. If these
groups are not abelian or $p \neq 2$, they have the following
special forms
\[
Q_1^1=<x,T| x^p=1;\, x^{-1}Tx =T^{1+p^{m-1}}>,
\]
and
\[
S=<y,T| y^p=1;\, y^{-1}Ty =T^{1+p^{m-1}}>.
\]
However, the conditions on $x$ and $y$ give that
\[
x^{-1}Tx= y^{-1}Ty \iff Tyx^{-1}=yx^{-1}T,
\]
so the only elements of $S_{p^m}$ which commute with $T$ (a complete
cycle of length $p^m$), are the powers of $T$. Thus $yx^{-1}=T^j$
for some $j$. Since the order of $yx^{-1}$ is $p$, the only choices
for $j$ are $j=p^m$ or $j=p^{m-1}$. For both choices we get
$S=Q_1^1$, namely $j=p^m$ gives that $x=y^{-1}$ (so $S=Q_1^1$), and
$j=p^{m-1}$ gives that $x=T^{-p^{m-1}}y$. Thus we have $x \in
<y,T>$, so that $<x,T>=<y,T>$, and hence $S=Q_1^1$. \qed

\begin{thm}
\label{lem:Qgroup1} Let $G$ be a subgroup of $S_{p^m}$ and $P$ a
$p$-Sylow subgroup of $G$ of order $p^s$ such that $T \in P$. Then
the following holds:
\begin{enumerate}
\item[(a)] If $s=m, P=<T>$.
\item[(b)] If $m<s\leq p+m-1$, then we have $P=Q_1^{s-m}$.
\end{enumerate}
\end{thm}
\pf From Lemma~\ref{lem:groupe} we have that $m \leq s \leq p^{m-1}+
p^{m-2}+ \cdots 1$. For the case $m=s$, it is obvious that $P=<T>$.
Now, let $s$ be such that $m<s\leq p+m-1$. Hence $P$ contains a
$p$-subgroup $P'$ of order $p^{m+1}$. By Theorem~\ref{lem:unique},
$P'=Q_1^1$. Let $j \geq 1$ be the largest integer such that $Q_1^j
\leq P$. If $j=p-1$, by Lemma~\ref{lem:odd} we have that $|Q_1^{p-1}|=p^{p+m-1}$.
Hence the assumption $s\leq p+m-1$ leads to
the unique solution $P=Q _1^{p-1}$. Thus, assume that $1\leq j
<p-1$. If $Q_1^j \neq P$, $N_P(Q_1^j)$ properly contains $Q_1^j$ and
by Lemma~\ref{lem:odd}, $N_P(Q_1^j)\leq Q_1^{j+1}$. As $Q_1^j \leq
N_P(Q_1^j)\leq Q_1^{j+1}$ and $Q_1^{j}\neq N_P(Q_1^j)$, by
Lemma~\ref{lem:2.6} $N_p(Q_1^j)=Q_1^{j+1}$, a contradiction of the
choice of $j$. \qed
\begin{thm}
\label{lem:Qgroup1} Let $p$ be an odd prime, $q=r^{\alpha}$ a prime
power, $C$ a cyclic code over $\mathbb{F}_q $ of length $p^m$, $m>1$
and $P$ a $p$-Sylow subgroup of $Aut(\mathcal{C})$ of order $p^s$
such that $T \in P$. Then the following holds:
\begin{enumerate}
\item[(a)] If $p \nmid \alpha$ and $p \nmid (d,r^a-1)$, then $s=m$, and
$P=<T>$ if and only if $Aut(\mathcal{C})=P\Gamma L(d,r^a)$, $d \ge 3$.
\item[(b)] If $p \geq 5$, $\alpha=1$ and $r=p$, $m>1$,
then $Aut(\mathcal{C})$ is an imprimitive group and $P$ is normal of
order $p^s$, $s>m$. If $m<s\leq p+m-1$, then we have $P=Q_1^{s-m}$.
 \item[(c)] If $z=1$, $p\nmid \alpha$ and $p \nmid (d,r^a-1)$, then $Aut(\mathcal{C})$ is an
imprimitive group and $P$ is normal of order $p^s \ge p ^{2m-1}$.
Furthermore, if $2m-1<s\leq p+m-1$, then we have $P=Q_1^{s-m}$.
\end{enumerate}
\end{thm}
\pf Statement (a) and the first parts of (b) and (c) follow from
Theorem~\ref{th:Zsigmondy}. We need only prove that if $s\le p+m-1$,
then $P=Q_1^{s-m}$. Assume $s \le p+m-1$, so that $P$ contains a $p$
subgroup $P'$ of order $p^{m+1}$. By Theorem~\ref{lem:unique}, we
obtain $P'=Q_1^1$. Let $j \geq 1$ be the largest integer such that
$Q_1^j \leq P$. If $j=p-1$, by Lemma~\ref{lem:odd} we have that
$|Q_1^{p-1}|=p^{p+m-1}$. Thus $Q_1^{p-1}$ is a subgroup of $P$ of
the same order as $P$, and hence $P=Q_1^{p-1}$, so we can assume
that $1\leq j <p-1$. If $Q_1^j \lneq P$, then $Q_1^j \lneq
N_P(Q_1^j)$ and by Lemma~\ref{lem:odd}, $N_P(Q_1^j)\leq Q_1^{j+1}$.
Since $Q_1^j \lneq N_P(Q_1^j)\leq Q_1^{j+1}$, by Lemma~\ref{lem:odd}
$N_p(Q_1^j)=Q_1^{j+1}$, which contradicts the choice of $j$. \qed
\begin{cor}
\label{lem:Qgroup} Let $\mathcal{C}$ and $\mathcal{C'}$ be two
cyclic combinatorial objects on $p^m$ elements, and let $P$ be a
$p$-Sylow subgroup of $\mathcal{C}$ such that $T \in P$. If
$|P|=p^s$ and $ s \leq p+m-1$, then $\mathcal{C}$ and $\mathcal{C'}$
can be equivalent only under the permutation of the following
subgroups of $S_{p^m}$:
\begin{enumerate}
\item[(i)] $AG(p^m)$ if $s=m$;
\item[(ii)] $Q^{s-m+1}$ if $s> m$.
\end{enumerate}
\end{cor}
\pf The result follows from Lemma~\ref{lem:brand},
Theorem~\ref{lem:Qgroup1} and Lemma~\ref{lem:odd}.
\qed

\begin{rem}
\label{rem:multi} Since each affine transformation can be written as
the product of a power of $T$ and a multiplier, and since $T \in
Aut(\mathcal{C})$ the power of $T$ is absorbed in
$Aut(\mathcal{C})$. Hence the permutation given in part (i) of
Corollary~\ref{lem:Qgroup} is reduced to a multiplier.
\end{rem}

In order to solve the isomorphism problem for cyclic combinatorial
objects, we must know the $p$-Sylow subgroup of $Aut(\mathcal{C})$.
To determine this, consider the following polynomial permutations
$f_1=T$ and $f_i(x)=1+x+p^{m-1}(x^2+\ldots+x^i)$ for $2 \le i \le
p-2$.

\begin{cor}
\label{cor:49}
Let $G$ be a subgroup of $S_{p^m}$ with a $p$-Sylow subgroup $P$,
and let $I$ be the largest value of $i$ such that $f_i \in G$. If
$I<p-2$, then we have $P=Q_1^I$.
\end{cor}
\pf Assume that $I$ is the largest $i$ such that $f_i \in G$ and
$I<p-2$. Let $P$ be a $p$-Sylow subgroup of $G$ of order $s$.
Let $s$ be such that $I+m \le s <p+m-1$.
From Theorem~\ref{lem:Qgroup1}, we have that a
$p$-Sylow subgroup of any subgroup of $G \le S_{p^m}$ which contains
$T=f_1$ and has order $p^s$ with $m\le s\leq p+m-1$. Then we have
$P=T$ when $s=m$ or $P=Q_1^{s-m}$, so in this case $s-m=I$.
Now, if $s \le I+m \le p+m-1$, we have from Theorem~\ref{lem:Qgroup1} that
$P=Q_1^{s-m}$, so $Q_1^I \cap G \le Q_1^{s-m}$. The assumption on
$I$ gives $I=s-m$.

Assume now that $s >p+m-1$. Since $I <p-2$, we have that $s>p+m-1
>m+I$. We will prove that this case cannot occur. We have $T=f_1 \in
Q_1^1$. From Theorem~\ref{lem:unique}, $Q_1^1$ is the unique
subgroup of $S_{p^m}$ of order $p^{m+1}$ which contains $T$. Hence
$Q_1^1 \lneq P$. Since $Q_1^1 \lneq Q_1^2$, it must be that $Q_1^1
\lneq Q_1^2 \cap P \le Q^2$. Hence from~Lemma~\ref{lem:2.6} we
obtain $ Q_1^2 \cap P = Q_1^2 $ which gives $Q_1^2 \le P$. Using the
same approach for $ 2\le i \le I$, we obtain $Q_I \le P$. The
assumption on $s$ gives that $ Q_I \lneq P$. Hence $Q_1^I \lneq
Q_1^{I+1} \cap P \le Q^{I+1}$ ($Q^{I+1}$ can be considered since it
was assumed that $I<p-2$). Hence from~Lemma~\ref{lem:2.6} we obtain
$Q_1^{I+1} \cap P= Q_1^{I+1}$. This contradicts the assumption on
$I$. \qed

This corollary suggest the following algorithm for $I<p-2$.

\hfill \\
\noindent
\textbf{Algorithm B}: Let $p$ be an odd prime, and  $\mathcal{C}$ and
$\mathcal{C'}$ be two cyclic combinatorial objects from the same category.
Then the equivalence of $\mathcal{C}$ and $\mathcal{C'}$ can be
determined as follows.

\textbf{Step 1}: Find the order of the Sylow subgroup of
$Aut(\mathcal{C})$ as follows. Find the largest $I$ such that $f_I
\in Aut(\mathcal{C})$. Then $s=i+m$, and do Step 2.

\textbf{Step 2}, find $f \in  Q^{I+1}$ such that
 $\mathcal{C'}= f \mathcal{C}$.

\begin{rem}
To find the required $I$ in Algorithm B we can use (for example) a binary
search which requires checking at most $\lceil \log_2 (p-1)\rceil + 1$ of the $f_i$.
Furthermore, the cardinality of $Q^{I+1}$ is $(p-1)p^{2m+I-2}$.
\end{rem}
\begin{defi}
The cycle graph on $n$ vertices is the graph $\mathcal{C}_n$ with
vertex set $\{0,1\ldots,n-1\}$ and $i$ adjacent to $j$ if and
only if $j-i\equiv \pm 1$.
\end{defi}
\begin{cor}
\label{iso:graph} Two cycle graphs $\mathcal{C}_n$ and
$\mathcal{C'}_n$ on $p^m$ vertices can be isomorphic only by a
multiplier.
\end{cor}
\pf The automorphism group of a cycle graph on $n$ vertices has
order $2n$~\cite[Ex. 2 p. 30]{Godsil}. Hence if $n=p^m$, there a is a
unique Sylow subgroup of $Aut(\mathcal{C}_n)$ of order $p^m$.
Then the result follows from Corollary~\ref{lem:Qgroup} and
Remark~\ref{rem:multi}. \qed

\end{document}